# OpenCluster: A Flexible Distributed Computing Framework for Astronomical Data Processing

Shoulin WEI[1,2,3], Feng WANG[1,2,*], Hui DENG[2], Cuiying Liu[2], Bo LIANG[2], Wei DAI[1,2,3] , Ying Mei[1,2,3], Congming Shi[2],Yingbo Liu[4], and Jingping Wu[2]

1, Yunnan Observatories, Chinese Academy of Sciences, China; wangfeng@acm.org

2, Computer Technology Application Key Lab of Yunnan Province, Kunming University of Science and Technology, China

3, University of Chinese Academy of Sciences, China

4, Yunnan academy of scientific & technical information, China

**Abstract:** The volume of data generated by modern astronomical telescopes is extremely large and rapidly growing. However, current high-performance data processing architectures/frameworks are not well suited for astronomers because of their limitations and programming difficulties. In this paper, we therefore present OpenCluster, an open-source distributed computing framework to support rapidly developing high-performance processing pipelines of astronomical big data. We first detail the OpenCluster design principles and implementations and present the APIs facilitated by the framework. We then demonstrate a case in which OpenCluster is used to resolve complex data processing problems for developing a pipeline for the Mingantu Ultrawide Spectral Radioheliograph. Finally, we present our OpenCluster performance evaluation. Overall, OpenCluster provides not only high fault tolerance and simple programming interfaces, but also a flexible means of scaling up the number of interacting entities. OpenCluster thereby provides an easily integrated distributed computing framework for quickly developing a high-performance data processing system of astronomical telescopes and for significantly reducing software development expenses.

## 1 Introduction

An increasing number of high-performance, high-precision, high-resolution telescopes have been used for astronomical observations. In the last decade, the data obtained by these observations have exponentially increased. The Sloan Digital Sky Survey (SDSS) telescope, for example, produces approximately 200 GB of data every night, adding to a database that was approximately 50 TB in 2012 [1]. In addition, the Large Synoptic Survey Telescope (LSST) has a three-billion-pixel digital camera and produces 5 to 10 TB of data each night [1, 2]. As the number of large datasets obtained in this way has drastically increased, many challenging problems have developed that demand prompt and effective solutions. Among these solutions are efficient distributed algorithms and frameworks, which can help address the scalability and performance requirements required by modern astronomical data processing.

Message Passing Interface (MPI) and MapReduce are two of the most significant approaches in high-performance computing research. However, in general, MPI is rarely selected for developing real-time data processing systems because it does not provide standardized fault tolerance interfaces and semantics. Although extensive research [27, 28, 29] has been conducted in this area, few available tools exist to help parallel programmers enhance their applications with fault tolerance support. Moreover, the exploitation of MPI is impeded by difficulties in software development. The original existing serial algorithms or programs must be modified or recreated to achieve performance improvement [32].

The renowned MapReduce paradigm [3], on the other hand, has been attracting considerable interest and is currently considered the best selection of a framework for large-scale data processing. Although different research groups have developed some astronomical image processing techniques by leveraging MapReduce [11, 12], only a few astronomical data processing pipelines have been developed using the MapReduce framework.

Overall, the existing distributed processing platforms are not yet well suited for astronomy. On one hand, most big data frameworks are heavily dependent on Java or Scala, whereas scientists are more likely to be familiar with Python or C++. On the other hand, it is not easy to simply copy or reuse the mature methods and frameworks of big data processing from the Internet. This is because the number of computing nodes of an astronomical project is far smaller than that of the Internet. Moreover, it is difficult to guarantee the processing performance in a limited computing environment. Meanwhile, the procedures and requirements of astronomical data processing pipelines are likewise different. Many attributes, such as scalability, real-time processing, and fault tolerance, are strongly demanded by the pipeline design.

Based on the above considerations, the main aim of our study is the design and implementation of a flexible architectural framework referred to as OpenCluster, which provides a distributed processing solution for huge amounts of astronomical data over few nodes or a cluster. In this paper, we discuss the OpenCluster design principles and detail its implementations. We highlight a case study of a method of designing a data processing pipeline for a telescope by using OpenCluster.

The remainder of this paper is organized as follows. In Section 2, we briefly overview related works on the methods and architectures of distributed processing. In Section 3, we present the OpenCluster design and implementation details as well as key technologies. In Section 4, we briefly introduce APIs of OpenCluster. In Section 5, OpenCluster use cases are presented. Section 6 describes the experiments performed to evaluate various performance aspects of the proposed architecture. Section 7 gives the discussions about OpenCluster. Finally, Section 8 provides our conclusions and directions for future work.

## 2 Related Work

In recent years, a surge of research activity in developing frameworks for distributed computing applications has occurred. Actually, the idea of using corporate and personal computing resources for solving computing tasks appeared more than 30 years ago. More recently, approximately ten years ago, various organizations began to use systems such as MPI cluster and Map/Reduce on account of advancements in global and local networks.

The first widely used distributed computing technique was the “grid” which was proposed by Foster and Kesselman [5]. Grid computing is distinguished from conventional distributed computing by its focus on large-scale resource sharing, innovative applications, and, in some cases, high-performance orientations. Open Grid Services Architecture (OGSA) is known as the next generation of grid architecture. Coupled with web service technology, OGSA is the most outstanding extension of service-oriented architecture [6]. A renowned example of distributed computing is the SETI@home project (setiathome.berkeley.edu). The project employs the computers of volunteers to filter signals from the Search for Extraterrestrial Intelligence (SETI) radio telescopes to search for extraterrestrial intelligence [7,8]. However, SETI@home enables use of private computing resources because it was created to solve only easily decomposed tasks that can be divided into non-coherent pieces.

MPI is another de facto standard for modeling a parallel program on a distributed memory system. Currently, the two major open-source MPI implementation code bases are OpenMPI and MPICH2 [30]. In the astronomical field, the tools characterized by HPC were built based on MPI, such as BDMPI, which was developed by the author of [31], and Mechanic, which was developed by the author of Sonina et al [33]. MPI is also widely used for scientific image processing in heterogeneous systems [32].

In the last decade, the most ubiquitous framework has been Apache Hadoop, an open-source software framework for storage and large-scale processing of datasets on clusters of commodity hardware that can reliably scale to thousands of nodes and petabytes of data [10, 11]. Apache Hadoop implements the MapReduce computational paradigm. Currently, Apache Hadoop is not just a framework; it is an ecosystem comprised of MapReduce and Hadoop distributed file systems (HDFSs), as well as a few related projects (Hive [15], HBase [16], and Pig [17]). Wiley et al. presented a method of using Hadoop to implement a scalable image-processing pipeline for the SDSS imaging database. They additionally described an approach to adapting an image co-addition to the MapReduce framework [12]. Furthermore, Tapiador et al. presented a framework as a thin layer on top of Hadoop without addressing a specific interface to the lower-level distributed system implementation (Hadoop) [13]. In addition, Ekanayake et al. designed a high-energy physics data analysis framework that was embedded in the MapReduce system by means of wrappers (in the Map and Reduce phases) and external storage. They additionally designed CGL-MapReduce, a streaming-based MapReduce implementation [14].

In addition to Hadoop, the frameworks that fit the MapReduce paradigm, such as Spark [18], Flink [19], and streaming technologies, such as Storm [20], are also widely used for enterprise and big data applications. These tools are largely driven by Internet companies and are most readily applied to web and business data.

## 3 OpenCluster

### 3.1 Motivation

We confronted many challenges while developing a high-performance data processing system for Mingantu Ultrawide Spectral Radioheliograph (MUSER) with MPI, MapReduce, Spark, and other frameworks. We realized the existence of limitations in current frameworks despite their effectiveness in general information technology. Obviously, using an existing solution of distributed data processing is an obviously preferable scheme that would save time. However, as mentioned above, MapReduce models are not well suited to astronomical data processing because the data format must be processed. The raw data generated by the observational equipment completely differs from the Internet data in terms of the amount, format, and potential users. Most existing big data frameworks are primarily intended for text-based data. The frameworks are written in (or ported to) Java. Reading and writing binary file formats, which are commonly used in astronomical research, are not straightforward.

To solve the problems of massive astronomical data processing, it is necessary to develop a framework that is cost-effective and can be simply used for all data processing systems. The framework should be able to run on heterogeneous systems and support different operating systems. Considering that few programming experts exist for the communication model and data exchange in an astronomical project team, the prospective framework should be capable of providing high-level interfaces with programming ease and using modern interpreted scripting languages with object-oriented features.

### 3.2 Fundamental Model

The processing of scientific data is similar to the production of industrial products. In general, a product experiences several phases, such as manufacturing, packaging, and shipping, before it is brought to market. Fig. 1 presents a schematic of production. Many managers who supervise a large group of employees respectively control the product procedure. The managers assign the related production tasks to specific workers. When a production process ends, the product is transitioned to the next procedure. In production, several public services, such as printing, phone use, and providing of meals, should be provisioned to all employees.

To construct a distributed computing framework for massive data processing, the model of modern industry can be referenced. In fact, the data processing procedure is very similar to industrial production. First, the astronomical telescope produces raw observational data. These data should be processed in a sequence of steps.

OpenCluster was designed to provide a unified framework that can easily support large distributed computation in a computer cluster environment. A system based on OpenCluster can be developed in a straightforward and efficient manner. The overall objective of OpenCluster is to permit a collection of heterogeneous machines on a network to be viewed as a general purpose concurrent computation resource. OpenCluster provides the communication backend based on sockets. It enables them to run on heterogeneous, geographically dispersed machines.

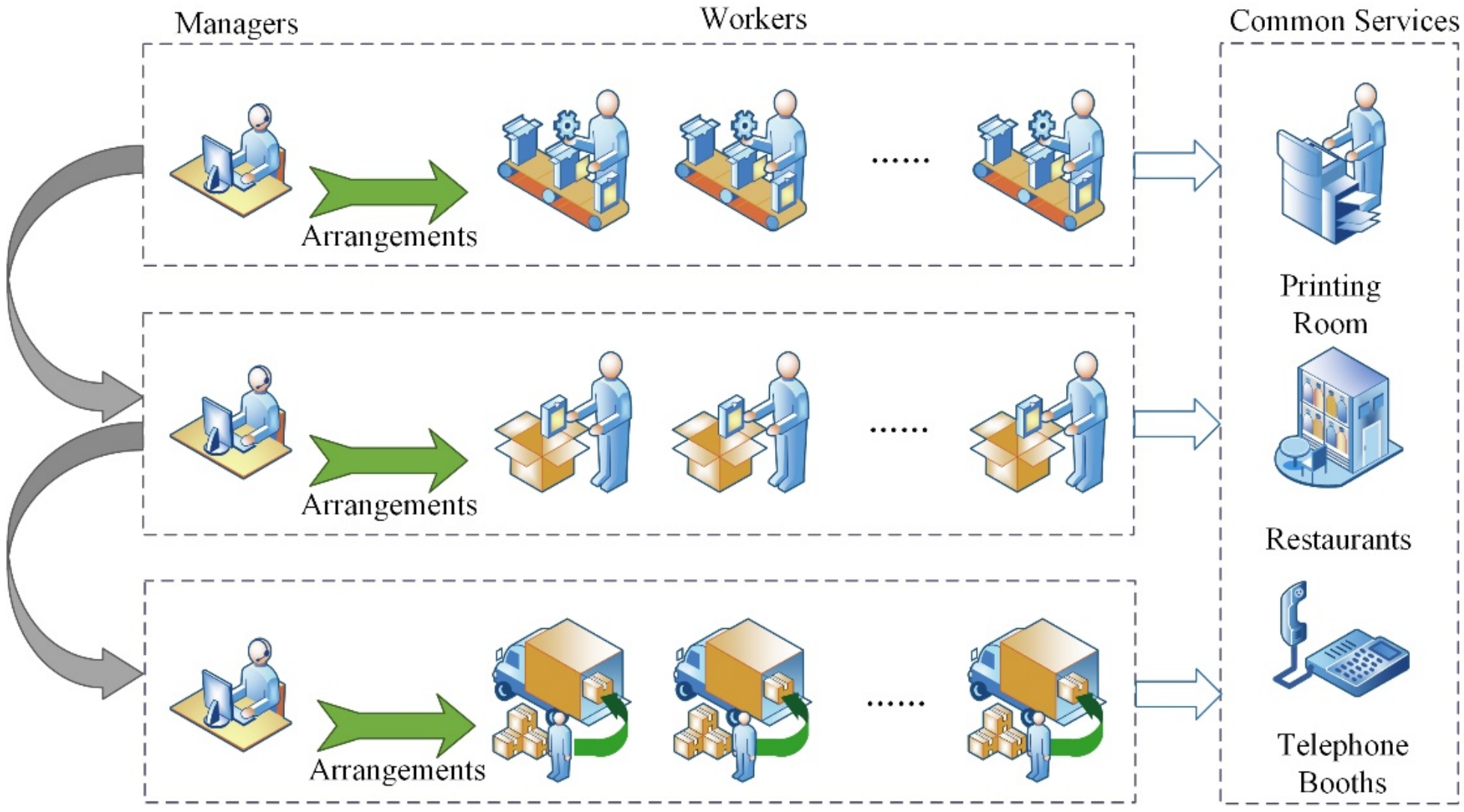

Fig. 1 Abstract process of traditional production.

OpenCluster was built in-house with minimal software requirements. Moreover, it does not depend on a particular middleware or analysis framework, thereby fostering greater flexibility in the installation. OpenCluster additionally includes a built-in monitoring system with no dependencies on external tools for this purpose. These properties make OpenCluster a very lightweight yet powerful tool and extend its scope beyond astronomical applications.

To devise an abstract model for OpenCluster, we adapt components from the modern industrial assembly line. Five main elements comprise the OpenCluster abstract model: factory, workshop, manager, worker, and service, (see Fig. 2).

- Factory: A logical registry that contains all information about workshops, managers, workers, and services. It adopts the master/slave model to avoid a single-point failure.
- Workshop: A computational node, which refers to a real computer in a common area. A daemon process runs in the workshop to communicate with the factory instance and to report information and status updates.
- Manager: A particular task or job manager that requests services and assigns tasks to workers.
- Worker: A process that runs in a workshop (computer). The workers wait for the call from managers.
- Service: A process similar to that of a worker. The service can be invoked by managers, workers, or even external applications.

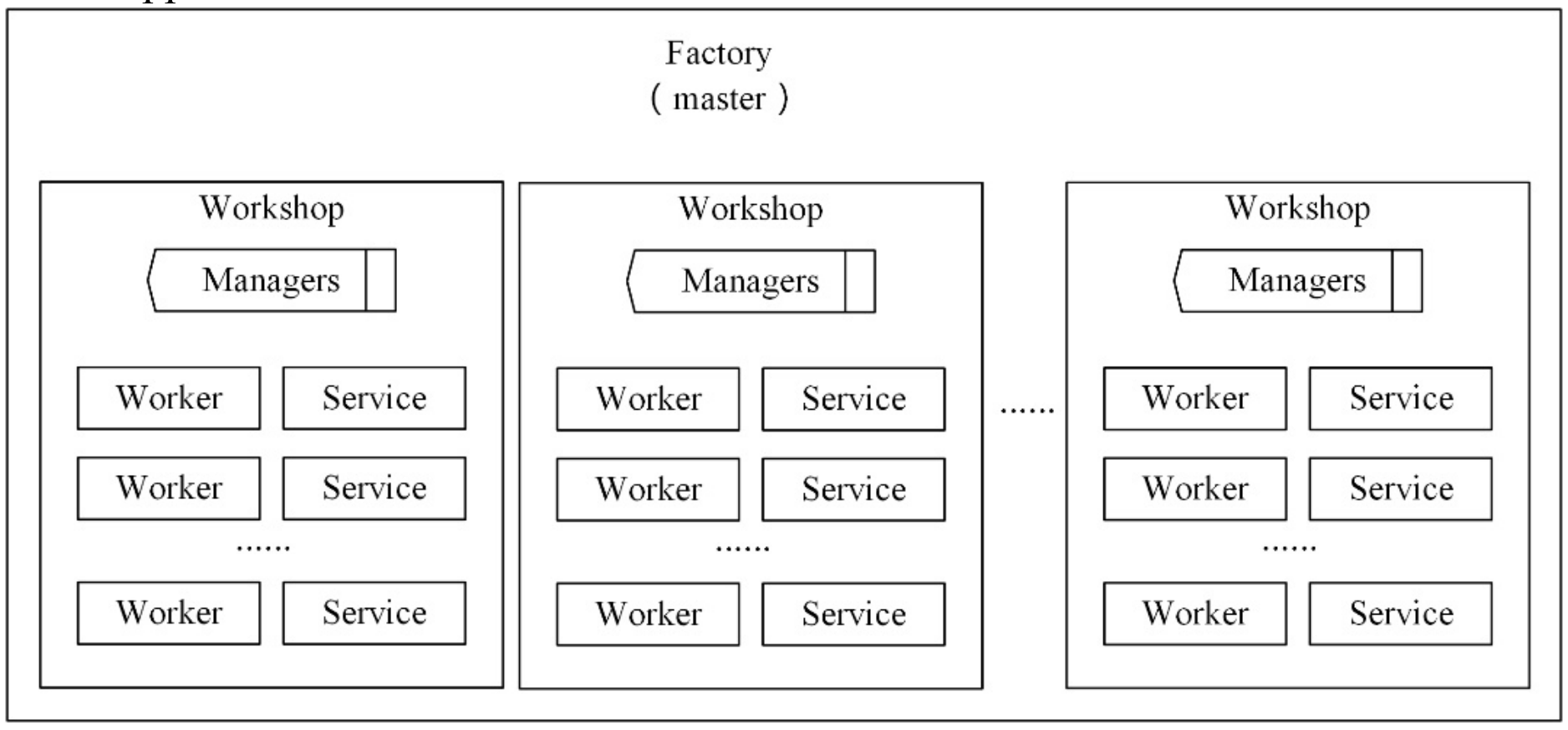

Fig. 2 Relational diagram of OpenCluster components.

## 3.3 OpenCluster Implementation

We ultimately selected Python to implement OpenCluster. Python is not effective for developing high-performance data processing software. Nevertheless, many astronomical scientists prefer Python on account of its simplicity and elegance. In addition, Python has a large user and developer base, and many scientific libraries are available for it. Most importantly, many emerging and mature programs exist in it for scientific computations written by astronomical scientists. These programs can be seamlessly and easily integrated into OpenCluster. To this end, we created five main base classes for implementing the functions of Factory, WorkShop, Manager, Worker, and Service. The other classes are derived from these five base classes.

### 3.3.1 Factory

The core component of OpenCluster, Factory administers one or more workshops. As a register center, all workshops, managers, workers, and services should first register to the factory. In general, we can launch at least two factory instances. One is a single master; the others are multi-slaves. These factory instances can run on a specified node or different physical nodes. If any connection information of the factory master changes, the factory master will synchronize the updated information to all slaves. Therefore, all slaves theoretically have the same information as the master. When a master factory failure occurs, the cluster selects one factory from slaves as the new master factory. It then immediately switches in real time to the factory. Thus, no single-point failure of the factory exists.

### 3.3.2 Workshop

We typically handle a real computing server in the cluster as a workshop. There are many workers or service instances running in the workshop. As a Linux daemon, Node Daemon runs in the background on each workshop of the cluster, respectively. It is responsible for managing workers, monitoring their resource usages (CPU, memory, and network), and reporting updates to the factory. The service interfaces provided by Node Daemon can be used to start or stop a new worker/service instance. On startup, Node Daemon registers to the factory and sends resource information of the node. A heart beat mechanism is used to monitor the connection between Node Daemon and the factory. If the factory does not receive the heartbeat, it can determine that problems occurred in the workshop. After a specific time period, the factory will remove the registration of the timeout Node Daemon. If Node Daemon does not receive the heartbeat, it will decide that problems exist with the current leader of the factories and a new leader is requested.

### 3.3.3 Worker

The worker is assigned to perform a task specified by the manager and returns the results to the manager. After the worker instance is successfully started, it attempts to connect to the leader factory. It then submits the information, including connection information, types of workers, and so on. After the information communication, the worker enters a listening state and awaits a call from the manager.

OpenCluster provides only a base class of workers (worker.py), which includes a function (“perform”) for executing tasks and setting attributes, such as the host, port, and work type (defined in the constructor). Workers defined by the user must inherit from the base class and override the function. The function *perform* has an argument with the *WorkPiece* type and returns a *WorkPiece* value. Section 3.3.5 discusses details of *WorkPiece*. In Section 4, we describe a method of extending a base class and implementing user-defined workers.

The set of methods defined by this base class include, but are not limited to, these functions. Additional functions also exist, such as the interaction with the factory, the heartbeat, and communication with managers, which all are hidden within the higher hierarchy of the base worker class. The user is not concerned with the part of the functions. Each worker instance is executed as an isolated process and also a socket server. This enables multiple worker processes to run in one node. However, the number of processes created in one node should work in harmony with the CPU core number. If we establish a greater number of processes than the core number, it will only reduce the processing speed.

#### 3.3.4 Manager

The manager is responsible for creating a job to be run, querying the factory, and expecting the factory to provide its workers or services with metadata containing the endpoint address and port of the provider matching the query. It additionally finds a series of suitable workers that can run the job and dispatch the job to these workers. When the manager is started, it looks up the list of workers who have registered in the factory and dispatches the job to the available workers. The workers then accept this request, perform the computing tasks, and return the results to the manager. The entire task can contain multiple managers. Like the worker, the scheduling logic of the manager is programmed by developers. Developers can freely control the scheduling process and assume different strategies. The manager is considered an entry to the computing tasks or an initial stage of a chain process.

The manager can invoke worker instances in an asynchronous or synchronous way. OpenCluster provides a lightweight thread pool implementation by which users can easily implement multi-threading. It depends on the ability of the computing node in which the manager resides to maximize the reuse of multi-core CPUs. OpenCluster additionally provides a manager base class (manager.py), which includes the methods for scheduling tasks (a function named “schedule” (task)). User-defined managers must inherit from the base class and override the function of the schedule. The function of the schedule includes only the argument of type *WorkPiece*. When it returns, the task is finished, and the manager will notify the factory of the corresponding task statistics.

#### 3.3.5 *WorkPiece*

To simplify programming, *DataItem* is added to OpenCluster, which inherits from the Python dictionary [22] and provides a set of convenient functions. *DataItem* can be regarded as an unordered set of key/value pairs indexed by keys, which can be any immutable types. Strings and numbers can both act as keys. The value can hold arbitrary data formats. Thus, we can send any data structures, such as text and binary data. This means that *DataItem* can accept any number or type of input parameters. It can also return any number or type of output results. Because OpenCluster is a framework, it should not assume the type and quantity of input and output parameters that are user-defined.

### 3.4 Deployment and Operation

OpenCluster is easy to deploy on a system. We only need to copy all files into a specific directory and set the related files with appropriate permissions. A web application was developed using the web.py [23] framework to control the starting and stopping of OpenCluster. It also can present the status of all operational nodes, managers, services, and workers. The status updates and statistics are reported by the clients to the factory. They provide useful information for monitoring the processing progress and for detecting errors. The updates include status changes and information about the execution host as well as manager statistics. The web server can run as a stand-alone daemon or as a CGI script within a more robust web server. The web interface additionally provides the functionality to control worker and service instances by authenticated users. Other features of the interface include graphs displaying resources usage and the number of managers in various states. Fig. 3 shows a screenshot of the OpenCluster interactive interface.

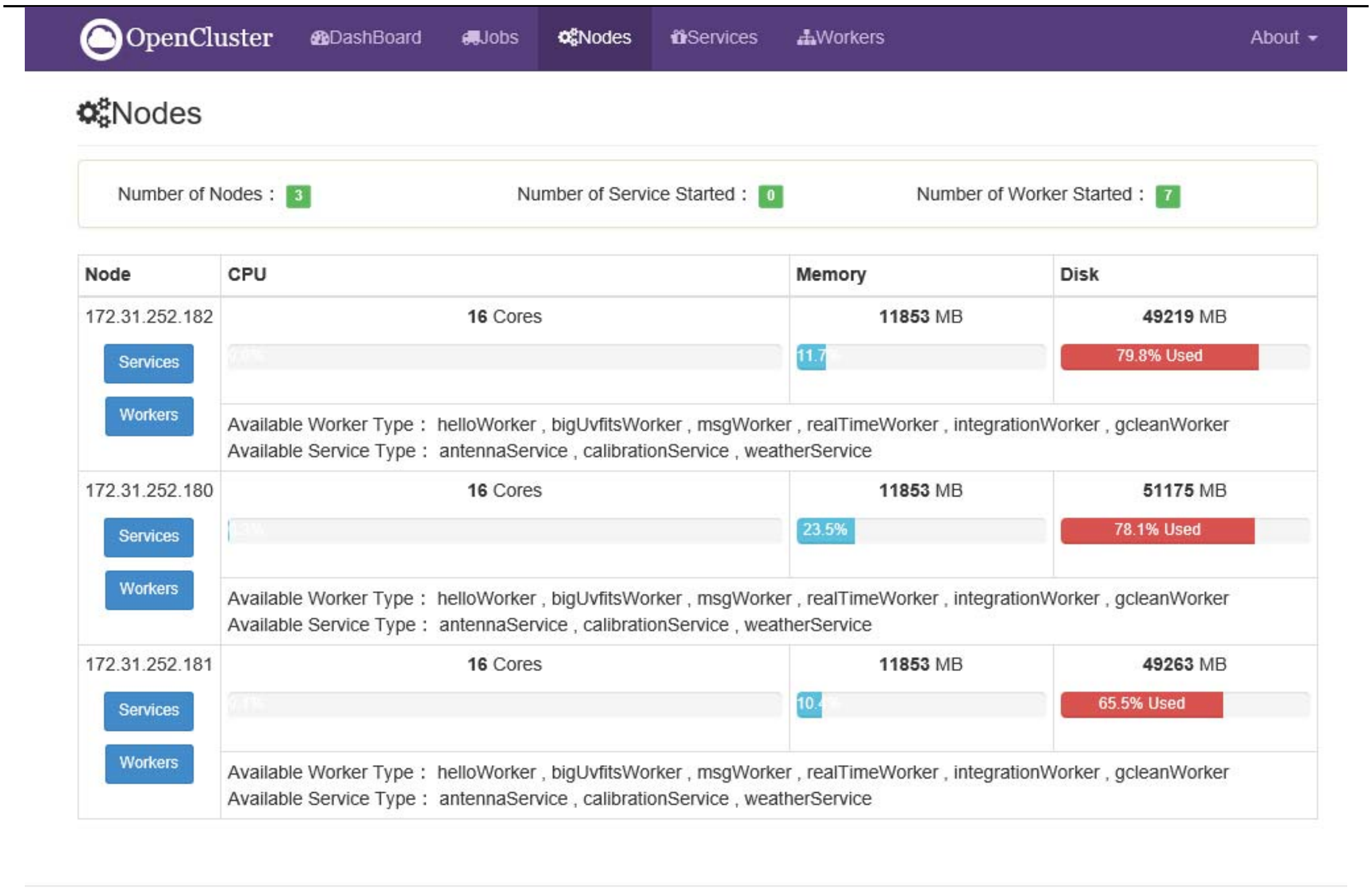

Fig. 3 OpenCluster monitoring interface.

Furthermore, OpenCluster has a variety of configurations for fine-tuning the behaviors of factories, managers, and workers. Each configuration has a default value defined in the file “config.ini,” where users can override these configurations. To run a cluster, users must first code the creation of a factory object and execute the factory *start* function.

## 4 OpenCluster APIs

We provide several APIs to help the user rapidly develop applications under OpenCluster. At a high level, each OpenCluster application consists of a manager program that runs the user’s main function, and several worker programs that receive tasks from managers in a cluster. The object-oriented method is adopted to implement a user-defined manager and worker. The complex codes, such as network communication, multi-threading, and task scheduling, are well wrapped so that users do not need to address them. To write an OpenCluster application, users must only implement the user-defined “managers” and “workers” by inheriting the base classes “Worker” and “Manager.” Both “Worker” and “Manager” include a required method (outlined in red in Fig. 4) to be overridden.

The base class “Manager” includes two methods:

- *getAvailableWorkers(worker_type)*, which returns available instances of the worker by a given type of worker in a cluster.
- *schedule(task),* which is an empty method that must be redefined in a derived class by users. In general, users split a task into sub-tasks, which they dispatch to available workers.

In the same way, the key methods of the class “Worker” are as follows:

- *ready(type,host,port)*. When invoked, it denotes that the worker is ready to accept the command. The first parameter is the type or identity of the worker; the second and third parameters are the host and port to which the worker instance binds.
- *perform(task)* is an empty method that users must override with detailed data processing logic. The parameter is “task” and the return result is a type of *WorkPiece*.

- *interrupt()* interrupts the running worker instance.

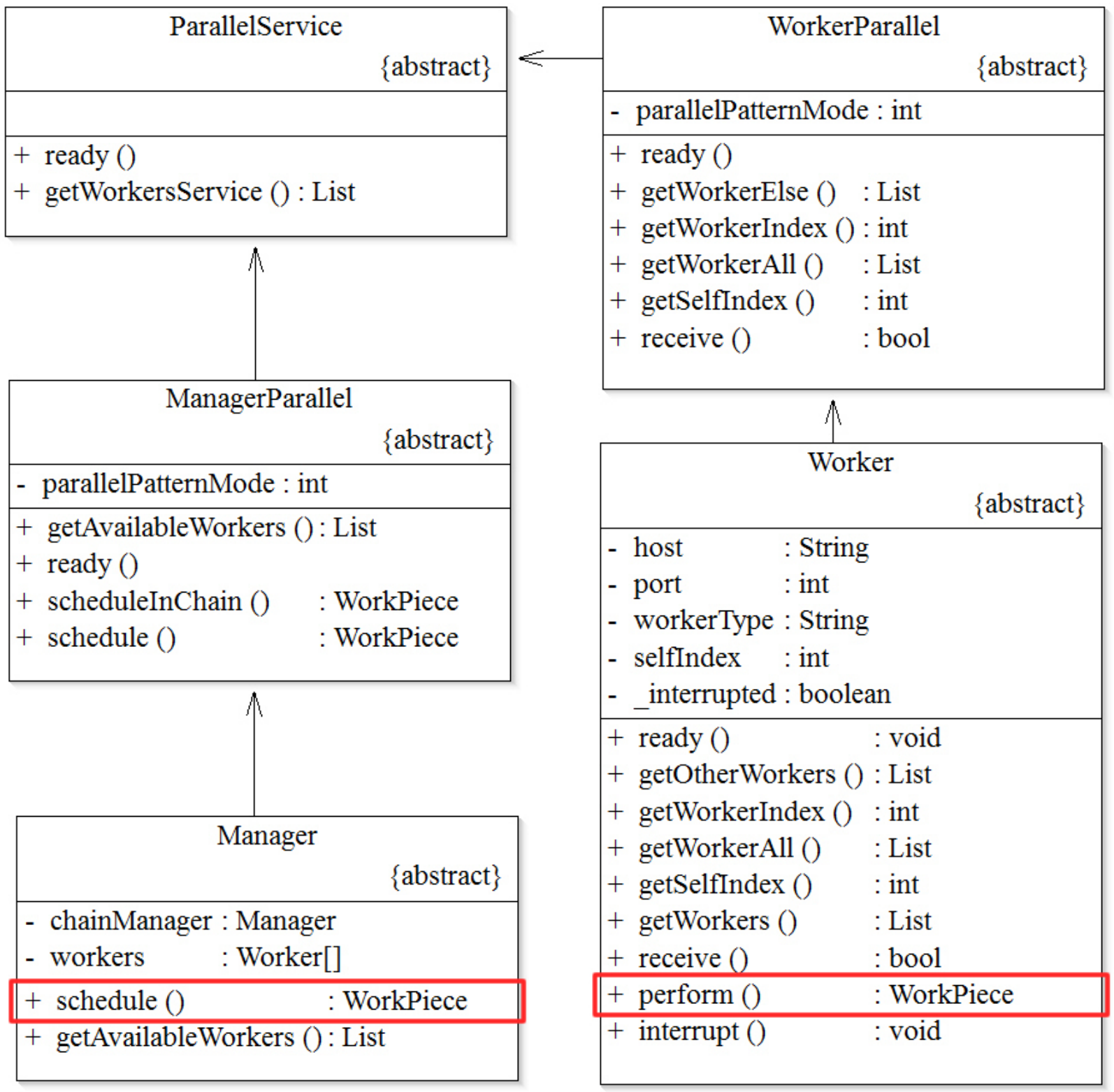

Fig. 4 Class hierarchy of “Manager” and “Worker.”

It is very easy to develop applications on the OpenCluster framework. The following pseudo-code (see Table 1) shows how to program a classic example “WordCount,” which is a simple application that counts the number of occurrences of each word in a given file.

Table 1. Pseudo code for Manager and Worker of WordCount in OpenCluster.

| | **Manager** | | **Worker** |
|---|---|---|---|
| 1 | **class WordCountManager(Manager) :** | 1 | **class WordCountWorker (Worker) :** |
| 2 | **def __init__** (self): | 2 | **def __init__**(self,name): |
| 3 | super(WordCountManager,self).__init__() | 3 | super(WordCountWorker,self).__init__() |
| 4 | **def schedule**(self, task): | 4 | self.name = name |
| 5 | workers = | 5 | **def perform** (self, work): |
| 6 | self.getAvailableWorkers("WordCount") | 6 | filepath = work.getObj("filepath") |
| 7 | wordcount = {} | 7 | offset = work.getObj("offset") |
| 8 | *# split task into sub tasks* | 8 | size = work.getObj("size") |
| 9 | for worker in workers : | 9 | wordcount = {} |
| 10 | subwordcount = worker.perform(*tasks[i]*) | 10 | *# count the number of words* |
| 11 | wordcount.join(subwordcount) | 11 | wh = WorkPiece(True) |
| 12 | **return** wordcount | 12 | wh.setObj("word",wordcount) |
| 13 | if __name__ == "__main__" : | 13 | **return** wh |
| 14 | manager = WordCountManager() | 14 | **if** __name__ == "__main__" : |
| 15 | result = manager.schedule("/logs/wordcount.log") | 15 | worker = WordCountWorker("worker1") |
| 16 | print result | 16 | worker.ready("WordCount","*",9280) |

Class WordCountManager is derived from the base class of Manager. The method “schedule” (lines 4 to 11) is overridden to split the task, dispatch sub-tasks to workers in turn, combine a result from the worker into the entire “wordcount” (line 10), and finally return the result. Class WordCountWorker inherits from the base class of Worker. It overrides the *perform* function (lines 5 to 13), where it receives the work from WordCountManager. It then opens the file, splits the line into tokens separated by whitespaces, and returns a work piece containing a key-value pair of <<word>, count>. In the main entry (lines 15 to 16), a worker instance is declared. After invoking the *ready* function, the worker with type "WordCount" listens to 9280 on all network interfaces.

## 5 MUSER Pipeline Using OpenCluster

OpenCluster is adaptable for developing various distributed computing schemes. It ports an existing project into a distributed computing environment with minimal customization and modification. OpenCluster users must only be familiar with the basic usage and astronomical data processing of Python; they are not required to have professional skills or experience in distributed computing or multithreading.

The data processing pipeline is one of the most important parts of MUSER. The massive data generated by MUSER should be processed in real time. MUSER generates 1.92 GB data per minute (low frequency sub-array; MUSER-I ) and 3.6 GB per minute (high frequency sub-array; MUSER-II) [9]. The amount of raw data can be up to approximately 4.4 TB every day, assuming 10 h of observation occurs in one day.

To fully leverage the computing resource, we designed a pipeline that can concurrently support multiple run levels. An architecture diagram of the MUSER data processing pipeline is shown in Fig. 5. Owing to the limitations of the hardware platform, the MUSER data processing pipeline simultaneously supports two modes of data processing. One is a batch data processing mode, which is used for scientific research. The other is real-time data processing, which is used for observational monitoring in daily observations. The tasks of real-time data processing must be guaranteed to ensure daily observation. Meanwhile, other batch data processing tasks can be processed when the computer platform has spare resources.

We thus present two examples of designing data processing pipelines with multiple concurrent tasks using OpenCluster.

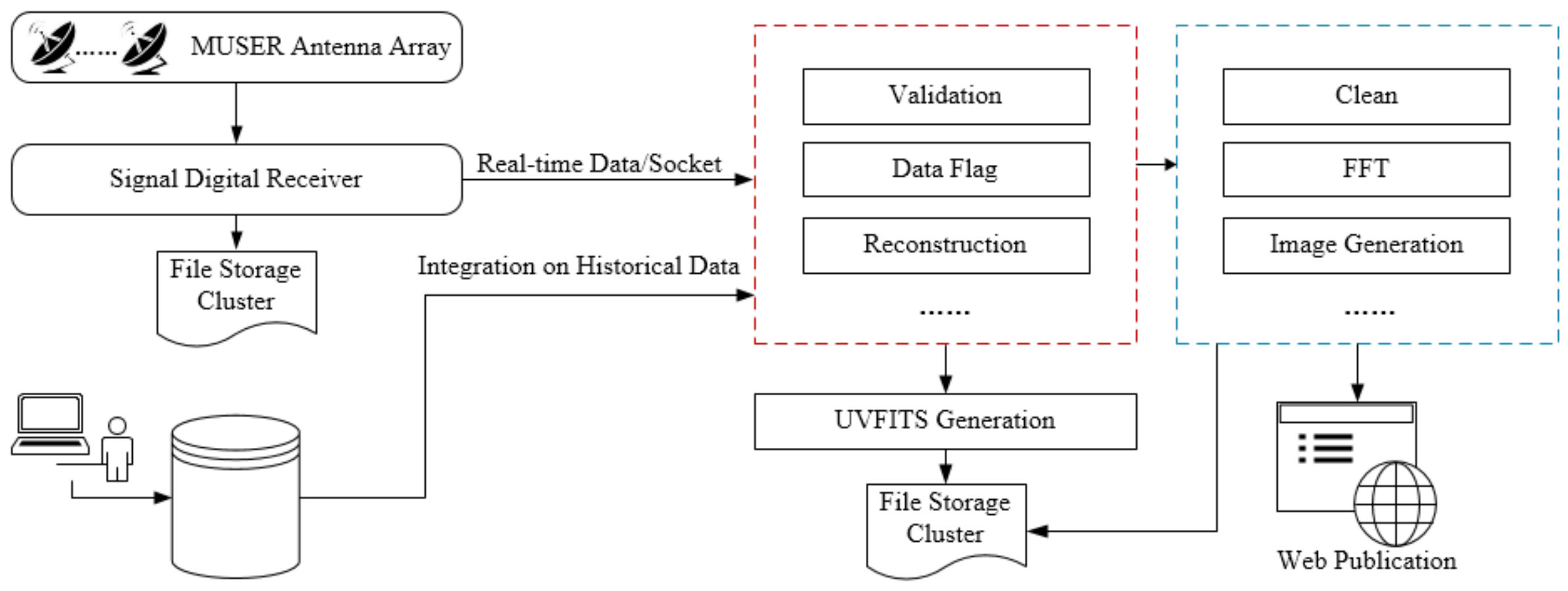

Fig. 5 MUSER data processing pipeline.

***Example 1. Data file format transformation***

The function of the file format transformation is necessary for follow-up scientific research and cooperation. However, the MUSER raw observational data cannot be directly shared because the raw data have no corresponding observation information, such as observational targets and target positions. To implement a function by OpenCluster to support converting raw data to a UVFITS file, we first developed a standalone function named *generateUVFITS*, which can open a specified file, read related raw data in a time range, and finally generate UVFITS files to the specified output directory. To integrate *generateUVFITS* into the OpenCluster platform, we simply create three classes:

1) UVFITSWorker, which is derived from the "Worker" class and wrapped *generateUVFITS* in the perform() method;

2) UVFITSManager, which is derived from the "Manager" class. The schedule() method is implemented to

schedule UVFITSWorker;

3) UVFITSTaskDispatcher, which is a daemon program that periodically polls a task queue to determine if it has a new task to submit.

In the UVFITSTaskDispatcher implementation, we created a thread pool of UVFITSManager (the optimum pool size should be equal to the number of CPU cores) to improve the computing performance. When a new batch computing task is submitted, this task is stored in a task queue. If there is an available thread in the thread pool of UVFITSManager, UVFITSTaskDispatcher presents a UVFITSManager instance from the thread pool. It then begins to run a new thread. When the UVFITSManager thread is started, the corresponding task is split into a series of sub-tasks. Then, the subtask is processed by the corresponding UVFITSWorker, respectively and concurrently. The flow diagram of data file format transformation is shown in Fig. 6.

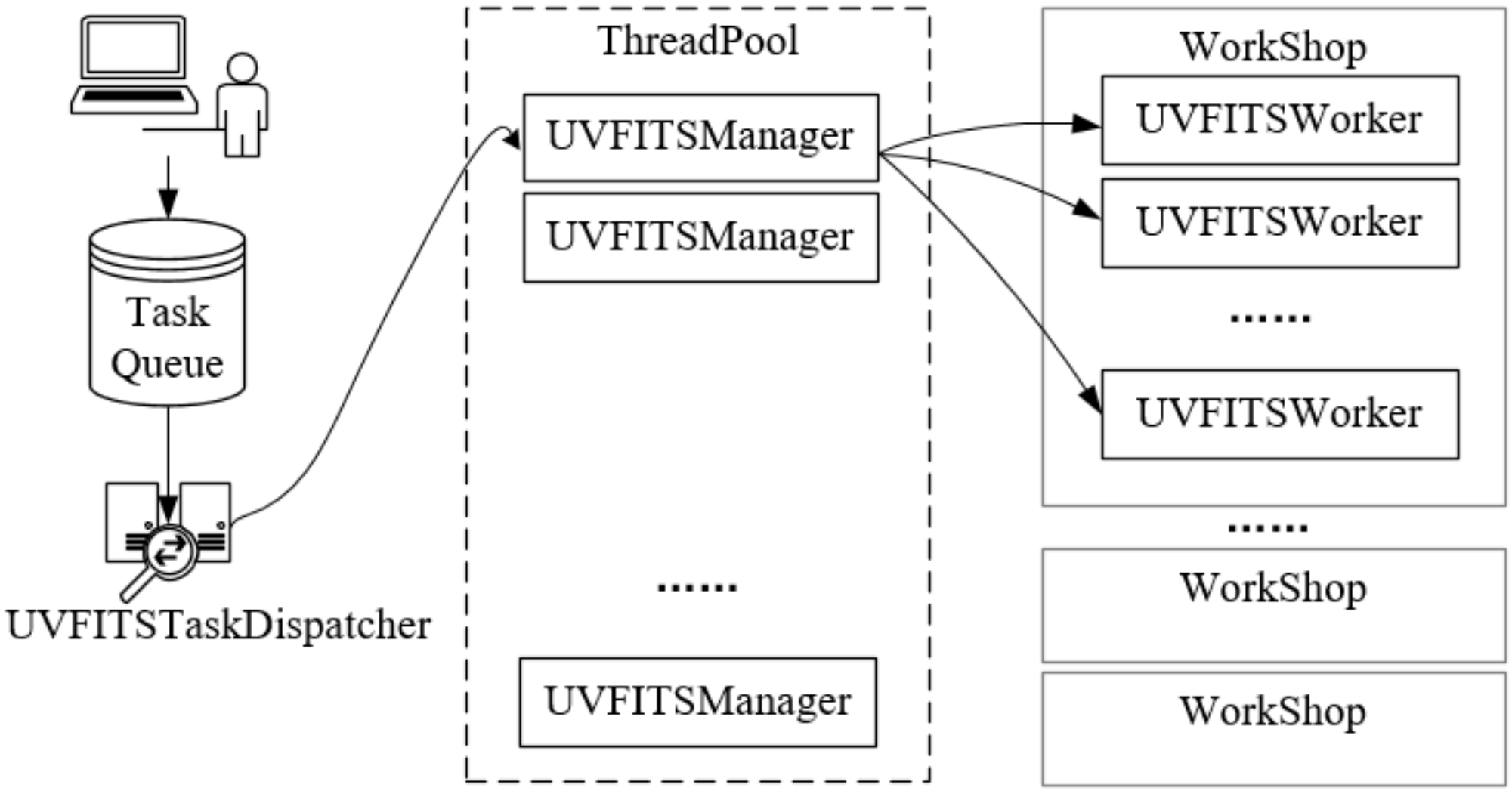

Fig. 6 Flow diagram of data file format transformation.

### *Example 2. Real-time imaging and monitoring*

As a synthetic aperture radio interferometer, MUSER can ultimately produce high-resolution images after performing a series of processes, such as gridding, flagging, and CLEAN [26]. Among them, the de-convolution for dirty images is a highly time-intensive calculation procedure that must be executed on machines with GPUs.

OpenCluster can design a complex data processing pipeline that can utilize hybrid hardware resources for MUSER. For example, observational data monitoring is a necessary function for daily observation. A dedicated computer will send raw observational data every five minutes to the monitoring system. The monitoring system should receive the data and imaging as quickly as possible. OpenCluster provides a TCP socket server implementation that can be reused by users to extend user-defined data handling in a pooling thread style.

1) MUSERSocketServer is an asynchronous TCP socket server. After new data is received, a new MUSERImageManager instance is invoked to process the new data.

2) MUSERImageManager, which is derived from the Manager class and rewritten as a *schedule*() method. The task is split into sub-tasks, and it arranges sub-tasks for MUSERImageWorker instances;

3) MUSERImageWorker, which is derived from the Worker class, is deployed on machines with GPUs. The business logics are implemented in the *perform*() method to fetch sub-tasks, perform computing issues (i.e., CLEAN, FFT, and so on ), and return results.

When MUSERSocketServer receives raw observational data, it creates a new MUSERImageManager instance and stores the instance in a thread pool queue. When resources are available to be run, the instance of MUSERImageManager is presented from the queue and executed. MUSERImageManager splits one observational data item into multiple slices according to the band and frequency. It dispatches each slice to multiple MUSERImageWorker instances to respectively generate images.

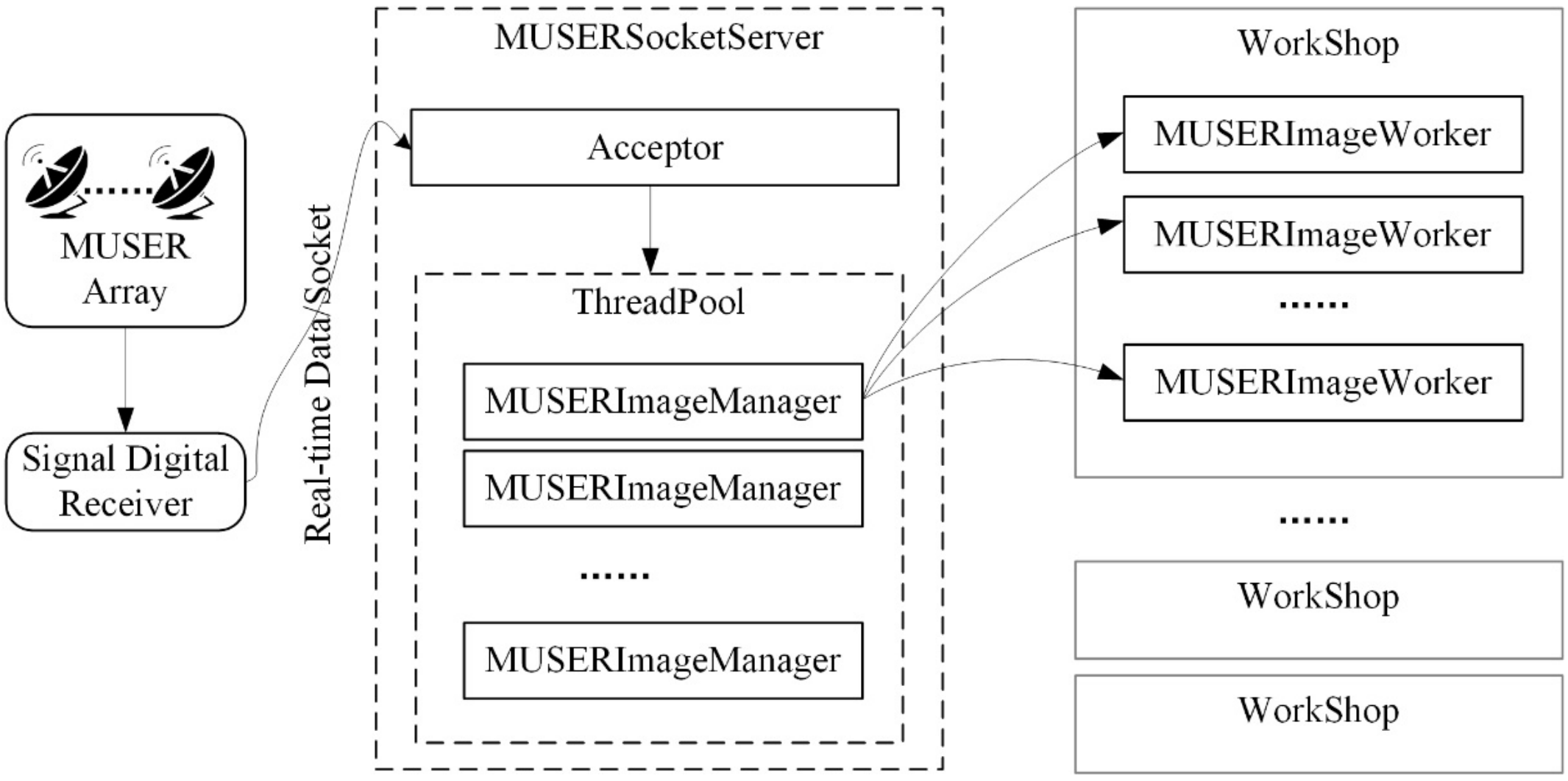

Fig. 7 Flow diagram of real-time imaging.

## 6 Performance Assessment

We conducted an assessment of OpenCluster because it is fully developed by Python and some developers may question its performance. We thus conducted a series of experiments to specifically assess its availability and performance. The testing environment was a computing cluster with five machines connected by gigabit Ethernet. Each cluster node had two-way Intel Xeon E5-2640 v2 CPUs, 2.0 GHz, 16 cores, and 1 TB of hard disk space. They ran runs on the CentOS 7 operating system. In addition, we used two GPU-enable nodes equipped with NVIDIA Tesla C2060. We exclusively assigned one node as the master factory, which also provided web monitoring, and other nodes as workers.

### 6.1 Message/Data transmission

In the first test, we compared the performances of MPI and OpenCluster by evaluating the message/data transmission (e.g., binary data transmission). Five nodes were used, and 16 worker instances were started on each node while assessing the OpenCluster performance. To compare it with MPI (programming with MPICH) [26], we established five nodes and created 16 processes on each node, respectively. The network environment of two platforms was the same.

We benchmarked the time consumption when sending a certain size of blocks from one node to others (or from one worker to other workers). The performance with a block size of 100 KB is shown in Fig. 8(a); Fig. 8(b) shows the results using 5 MB. By comparing the performance results between 100 K and 5 M of the block size, we determined that OpenCluster performed much worse than MPI. However, the performance became much closer to MPI when using a large block. This result implied that remote parallel execution was not favorable for smaller tasks because the overhead for communication between the machines was significant compared to MPI.

### 6.2 Execution performance

In the second test, with the same five nodes and a split raw file containing 1,000 frames, we evaluated the time used during UVFITS generation in MUSER for consecutive increasing of the quantity of worker instances (as shown in Fig. 9). The existing code of UVFITS generation could only be executed on a standalone machine without multi-threading, and the execution time for one frame was approximately 0.3 s. With a massive quantity of frames, it was a very time-intensive procedure. However, with the extensibility provided by OpenCluster, the same code could be executed on five nodes, and each node contained 16 instances. With an increased quantity of worker instances, the generation cost linearly decreased. We could thus conclude that the calculation could be performed in parallel, thus reducing the total computing time.

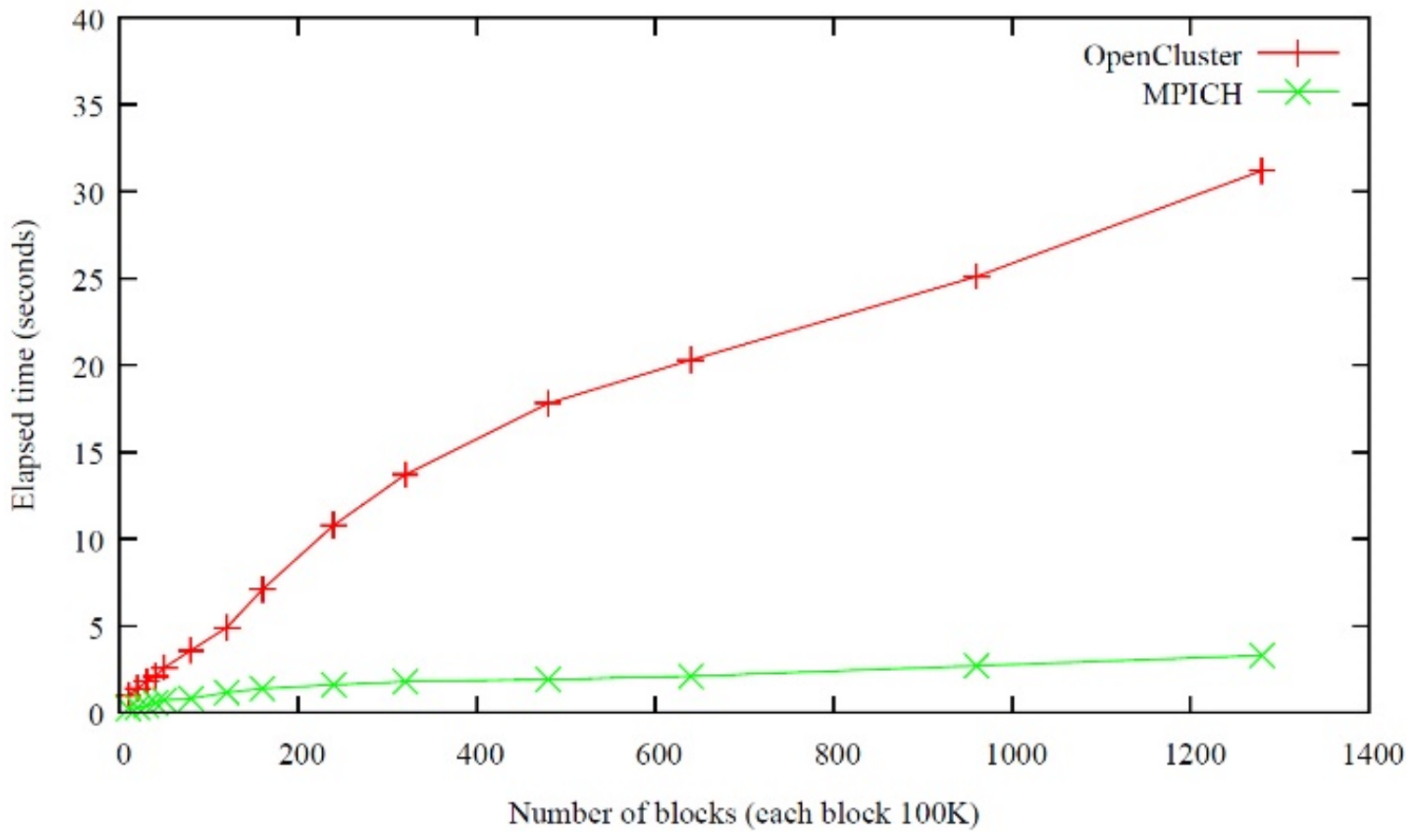

Fig. 8(a) Comparison of elapsed times for binary transmission using a 100-K size block between OpenCluster and MPI.

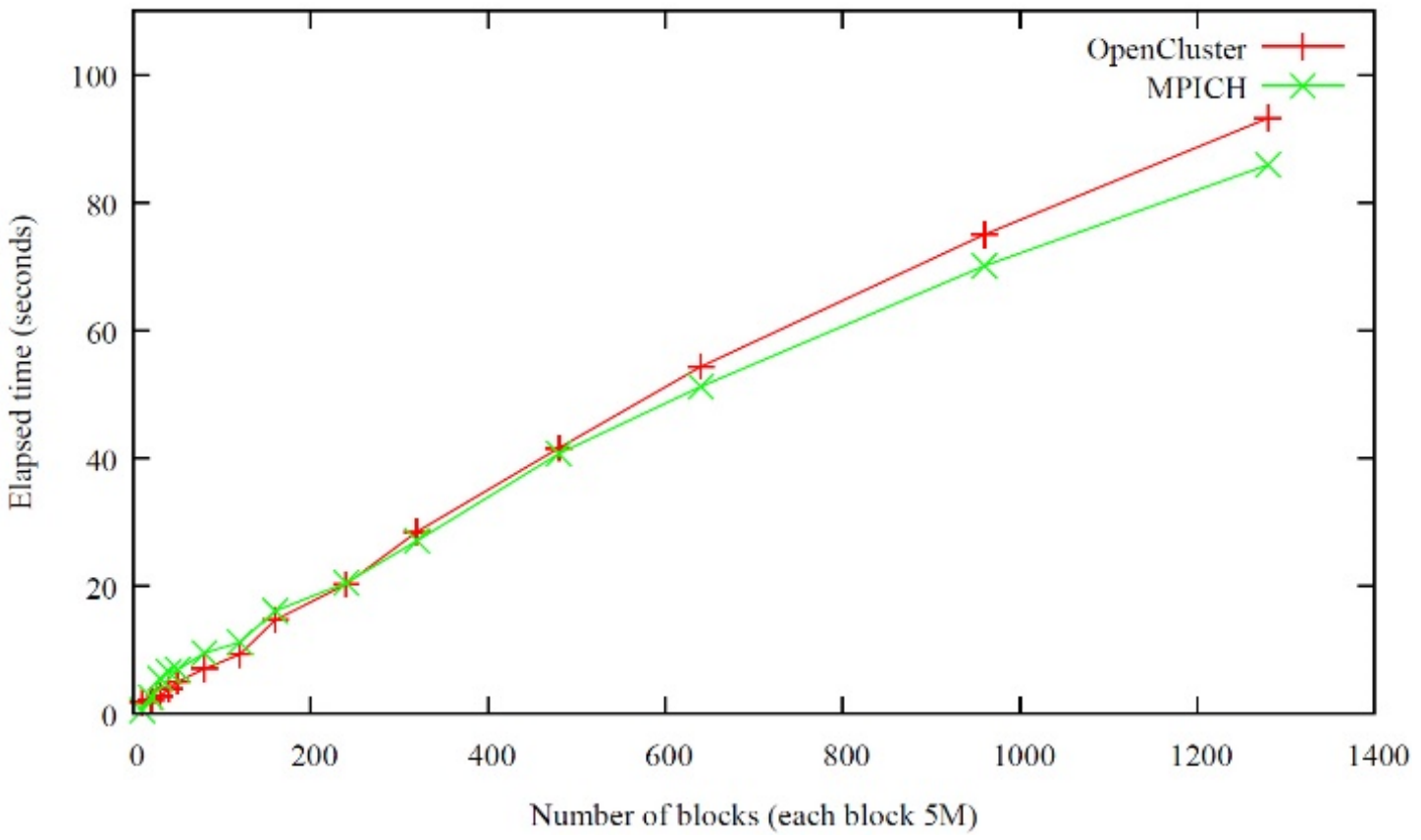

Fig. 8(b) Comparison of elapsed times for binary transmission using a 5-M size block between OpenCluster and MPI.

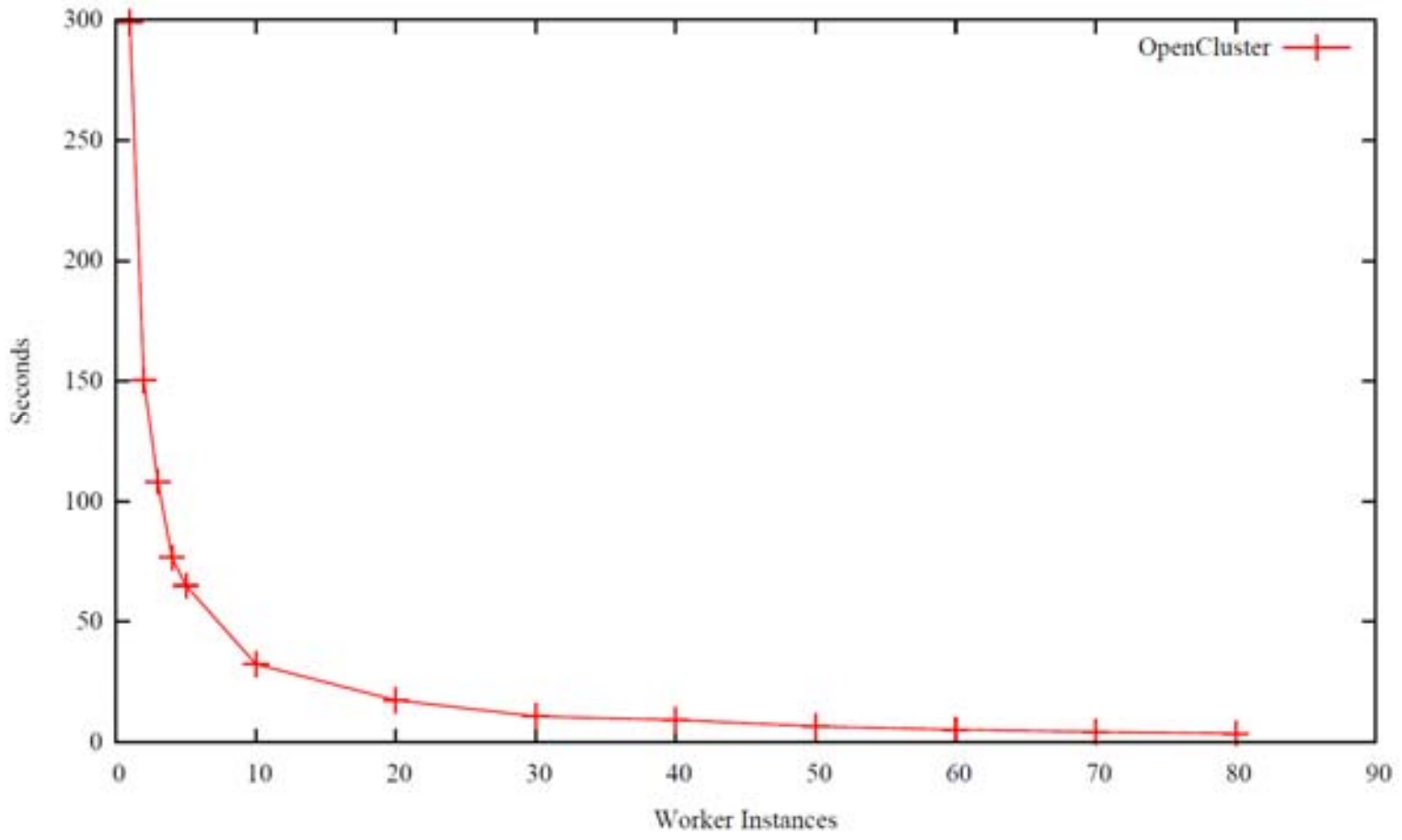

Fig. 9 Elapsed times for UVFITS generation evaluation of 1,000 frames for different numbers of worker instances.

According to the above results, with OpenCluster, the MUSER data processing procedure was spread across multiple computers and a GPU hybrid system. This made it easy to achieve optimal parallelism and maximize the usage of the cluster. It additionally provided an acceptable level of performance to meet the MUSER requirements. Moreover, its advanced scale-out capability, as demonstrated, will enabled customers to rapidly expand an infrastructure with minimal efforts.

## 7 Discussion

### 7.1 Stability

We considered several mechanisms to enhance OpenCluster stability. On account of distributed system constraints, at least one node should act as a leader responsible for cluster management and scheduling. Therefore, when the cluster is operating at various stages, it is necessary to ensure that a leader exists to provide services in the event of malfunctions. If a failure occurs, the system must be able to select a new leader.

The method of leader selection in OpenCluster is different from that of Paxos proposed by Lamport et al. [21]. Paxos is a protocol of processes for reaching a consensus on a series of proposals. In terms of maintaining the consistency of leader selection or variable modifications, Paxos adopts a considerable majority approval mechanism, similar to a parliamentary vote. For example, leader selection is regarded as a bill that requires more than half of the voter affirmations. Each node has a record number after carrying the bill. When the node again receives a request of another leader candidate, it rejects the request because it already has the record number.

Nevertheless, a lightweight courtesy method is adopted in leader selection of factories in OpenCluster. When a factory instance is started, it asks other factory instances if they would like to be a leader. If there is no leader, the factory performs as the leader. If a factory was already the leader, the factory will perform as a slave. The benefit of this courtesy approach with no competition possibly avoids conflict and maintains a consistency between factories. Fig. 10(a) shows the process of leader selection of a factory.

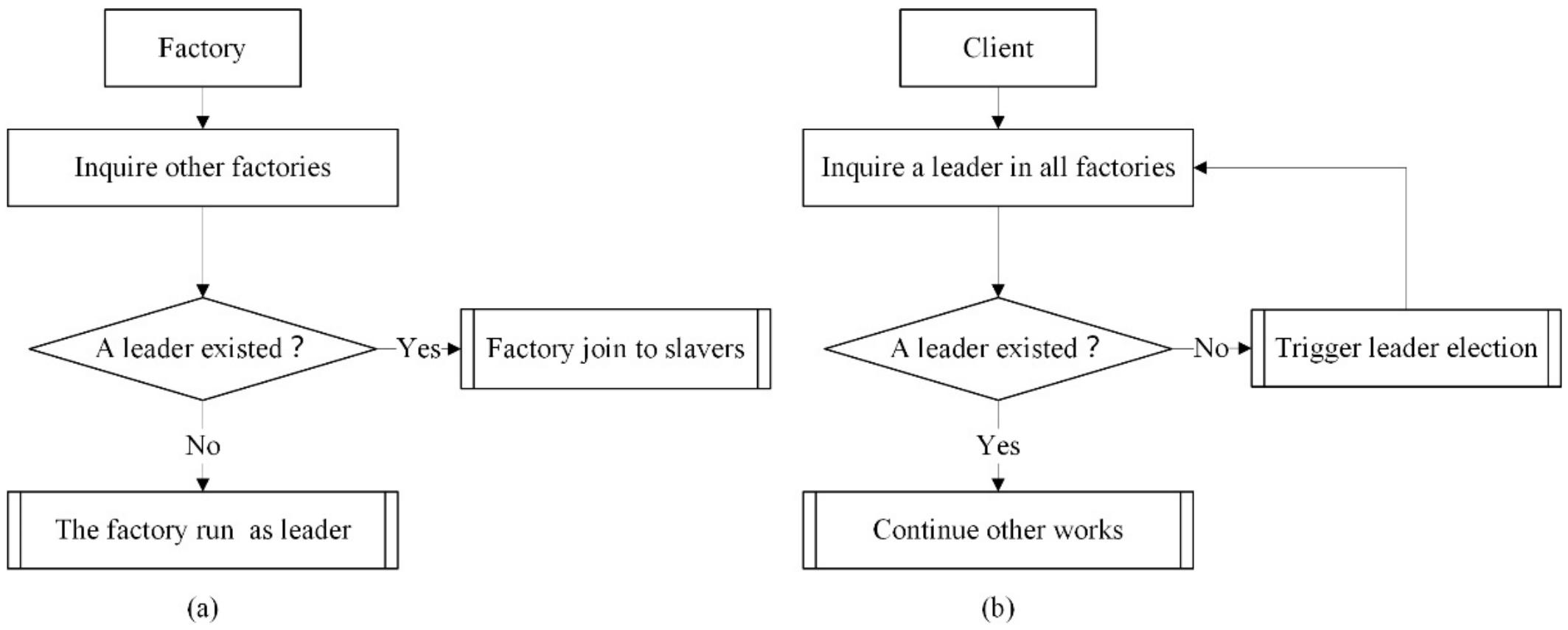

Fig. 10(a) Process of leader selection of a factory; (b) client operational processes before other works.

The leader automatically duplicates updates to other factory instances via a background thread. A factory is responsible for providing a registration and lookup for connections of workshops, workers, services, and managers (categorized as clients), just like a directory service, whereas it does not dispatch the computing tasks or data. Thus, much less stress occurs on the CPU or memory in a factory instance. Once the factory leader is determined, the clients deal with the leader. Therefore, the clients must know who the leader is. Fig. 10(b) illustrates the client operational processes before detailed works.

### 7.2 Hybrid Resource Scheduling

OpenCluster can utilize hybrid hardware resources on a computing platform that has CPU-only and CPU/GPU hybrid computers. With the OpenCluster schedule, CPU-oriented tasks can be executed on general-purpose nodes (all computers), while time-consuming image processing tasks can be executed on GPU-enabled machines (e.g., CPUs/GPUs). This feature is quite useful for designing high-performance pipelines, especially pipelines locally executed in an observatory. It is very difficult to implement high-performance imaging tasks without GPU support. However, on account of the power consumption constraint, only parts of computers are equipped with professional GPU cards.

OpenCluster supports hybrid resource scheduling mechanisms because of its registration approach of workers. During startup, the worker must register to the factory the type of task that it can accomplish. Therefore, managers can differentiate varieties of workers and dispatch different tasks to relevant workers that run on the appropriate hardware.

## 8 Conclusion

1, We presented a novel distributed framework, i.e., OpenCluster, for computing in astrophysical field. The system has been implemented and released at https://github.com/astroitlab/opencluster. The data processing pipeline of MUSER which designed upon OpenCluster proves that OpenCluster is robust, reliable and scalable. We presented the concept, the design, the programming interfaces and the components of OpenCluster.

2, We have demonstrated applications of the OpenCluster framework that show different aspects of the usage of the framework. We also described an automated way that the OpenCluster provides to monitor the behavior of each task and take actions, based upon the observed behavior.

3, The current OpenCluster design has some limitations, which must be considered when using the framework. OpenCluster is currently being enhanced to support processing based on the messaging center and self-defined task prioritization. Moreover, it does not support resource isolation. Thus, we are considering integrating it with Mesos [23]. We aim to provide a hierarchical cluster-based architecture that can be dynamically tuned to different workloads in the astrophysical field. In the future, we will apply the architecture in other scientific areas to verify its effectiveness and improve its efficiency.

The work discussed in this paper was jointly supported by Kunming University of Science and Technology and Yunnan Observatories of Chinese Academy of Sciences. This paper is funded by National Natural Science Foundation of China under Grants No.U1231205 and No.11403009, and National Natural Science of Yunnan Province under Grants No. 2013FA013, 2013FA032 and 2013FZ018. We also thank the reviewers for suggestions that improved the paper.